\documentclass[aps,pra,twocolumn,groupedaddress,amsmath,showpacs]{revtex4}
\usepackage{graphicx}
\usepackage{pst-node}
\DeclareGraphicsExtensions{.eps}
\newcommand{\ket}[2]{\mbox{$|#1\rangle_{#2}$}}
\newcommand{\bra}[2]{\mbox{$_{#2}\langle #1|$}}
\newcommand{\braket}[2]{\mbox{$\langle #1|#2\rangle$}}
\newcommand{\qcnot}{\ensuremath{\mbox{\textsf{CNot}}}}
\newcommand{\qgcp}[2]{\ensuremath{\mbox{\textsf{CP}}_{#1}\left(#2 \right)}}

\newcommand{\qgswap}[1]{\ensuremath{\mbox{\textsf{Swap}}_{#1}}}
\usepackage{amssymb}

\begin{document}

% Use the \preprint command to place your local institutional report
% number in the upper righthand corner of the title page in preprint mode.
% Multiple \preprint commands are allowed.
% Use the 'preprintnumbers' class option to override journal defaults
% to display numbers if necessary
%\preprint{}

\title{Selective recoupling and stochastic dynamical decoupling}

% repeat the \author .. \affiliation  etc. as needed
% \email, \thanks, \homepage, \altaffiliation all apply to the current
% author. Explanatory text should go in the []'s, actual e-mail
% address or url should go in the {}'s for \email and \homepage.
% Please use the appropriate macro foreach each type of information

% \affiliation command applies to all authors since the last
% \affiliation command. The \affiliation command should follow the
% other information
% \affiliation can be followed by \email, \homepage, \thanks as well.
\author{O.~Kern}
\email[e-mail: ]{oliver.kern@physik.tu-darmstadt.de}
%\homepage[]{Your web page}
%\thanks{}
%\altaffiliation{}
\author{G.~Alber}
\affiliation{Institut f\"ur Angewandte Physik, Technische Universit\"at Darmstadt, D-64289 Darmstadt, Germany}

%Collaboration name if desired (requires use of superscriptaddress
%option in \documentclass). \noaffiliation is required (may also be
%used with the \author command).
%\collaboration can be followed by \email, \homepage, \thanks as well.
%\collaboration{}
%\noaffiliation

\date{\today}

\begin{abstract}
An embedded selective recoupling method is proposed which is based on the idea of embedding the recently proposed
deterministic selective recoupling scheme of \citeauthor{YLMY04} \cite{YLMY04} into a stochastic dynamical decoupling method, such as the recently proposed Pauli-random-error-correction-(PAREC) scheme \cite{parec}.
The recoupling scheme enables the implementation of elementary quantum gates in a quantum information processor by partial suppression of the unwanted interactions.
The random dynamical decoupling method cancels a significant part of the residual interactions.
Thus the time scale of reliable quantum computation is increased significantly.
Numerical simulations are presented for a conditional two-qubit swap gate and for a complex iterative quantum algorithm.
\end{abstract}

% insert suggested PACS numbers in braces on next line
% insert suggested keywords - APS authors don't need to do this
%\keywords{}

%\maketitle must follow title, authors, abstract, \pacs, and \keywords
\maketitle

% body of paper here - Use proper section commands
% References should be done using the \cite, \ref, and \label commands
% Put \label in argument of \section for cross-referencing
\section{\label{sec:intro}Introduction}
Stabilizing quantum systems against uncontrolled perturbations is one of the major challenges in quantum information science.
In comparison with measurement-based quantum error correction, at first sight dynamical decoupling methods seem to suffer from
the disadvantage of not being capable of correcting unwanted perturbations perfectly. However, typically
measurement-based quantum error correction
requires ancillary quantum systems and complicated measurement and recovery operations \cite{nielsenchuang}
which are difficult to realize by nowadays technology. By contrast dynamical decoupling methods 
are able to suppress unwanted perturbations significantly even over long interaction times without
requiring additional ancillary quantum systems \cite{parec,VK05,combi,SV2005}.
Thus, in view of nowadays experimental difficulties in controlling many-particle quantum systems these latter methods
offer interesting perspectives, in particular, for the first generation of quantum information processors.

Recently a deterministic dynamical recoupling scheme
was proposed for dipole-coupled nuclear spins in a crystalline solid \cite{YLMY04}.
While in all previously proposed similar schemes \cite{JK99,LCYY00,SM01,Leu02} the evolution-time overhead
grows linearly with the number of spins, this particular scheme leads to an evolution-time overhead which is
independent of the number of spins involved.
Thus, it appears to be well suited for the stabilization of quantum information processors against unwanted inter-qubit interactions.
This recoupling scheme uses particular combinations of fast broadband and slower selective radio-frequency fields
to turn off all couplings except those between two particularly selected ensembles of spins.
Thereby, spins within each ensemble representing a particular logical qubit are decoupled \cite{LGYY02}.
Furthermore, cross-couplings between selected ensembles are avoided by requiring that qubit
couplings have to be much stronger than any other couplings within each ensemble.
Unwanted couplings are suppressed up to second-order average Hamiltonian theory
with the help of time-symmetric pulse sequences.
Despite many advantages in this recoupling scheme the residual higher-order interactions accumulate coherently
thus leading to a quadratic-in-time decay of the fidelity of any quantum state \cite{shep144}. This restricts 
the achievable time scales of reliable quantum computation significantly.

In this paper it is demonstrated that the performance of this recoupling scheme
can be improved significantly
by embedding it into a stochastic decoupling scheme. For this purpose 
the recently proposed Pauli-random-error-correction (PAREC)
method \cite{parec} turns out to be particularly useful. Thereby, the stochastic decoupling scheme
destroys the coherent accumulation of higher-order residual interactions to a large extent so that 
the fidelity decay of any quantum state is slowed down significantly to an almost linear-in-time one.
As a result reliable quantum computation can be performed on significantly longer time scales.

This paper is organized as follows:
The basic ideas underlying the recently proposed deterministic recoupling scheme of Ref. \cite{YLMY04}
are summarized briefly in Sec. \ref{sec:rec1} for the sake of completeness.
In Sec. \ref{sec:rec2} a simple restricted embedded decoupling scheme is introduced.
Though it already leads to first improvements in comparison with the deterministic selective recoupling scheme
of Ref. \cite{YLMY04},
its error suppressing properties
can still be improved significantly by an additional simple symmetrization procedure.
In Sec. \ref{sec:pi4gate} the stabilization properties of this symmetrized embedded recoupling scheme are analyzed
for a unitary conditional two-qubit swap gate.
In Sec. \ref{sec:saw} its stabilizing properties are investigated
by applying it to the iterated quantum algorithm of the quantum sawtooth map \cite{shep129}.

\section{Selective recoupling by embedded dynamical decoupling\label{sec:rec}}

\subsection{Deterministic selective recoupling of qubits\label{sec:rec1}}

In this section the basic ideas underlying the recently proposed 
recoupling scheme of Ref. \cite{YLMY04} are summarized. In particular, the form and magnitude of
the residual higher-order interaction is discussed which cannot be suppressed by the suggested %WHH-4 and Super-WHH
pulse sequences.

Let us consider $n_q$ nuclear spin-1/2 systems in a crystalline solid which are interacting with an external 
static magnetic field in $z$-direction. In the
rotating wave approximation
their  Hamiltonian is given by
\begin{equation}\label{eq:sysham}
\begin{split}
\hat{H} &= - \sum_{k=0}^{n_q-1} \frac{\hbar \omega_k}{2} \hat{\sigma}_z^k
+ \sum_{k<l=0}^{n_q-1} \frac{J_{kl}}{4}
\bigl( 2\hat{\sigma}_z^k\hat{\sigma}_z^l - \hat{\sigma}_x^k\hat{\sigma}_x^l - \hat{\sigma}_y^k\hat{\sigma}_y^l \bigr)\\
&\equiv \hat{H}_Z + \hat{H}_D
\end{split}
\end{equation}
with the Pauli spin operators $\hat{\sigma}_x$, $\hat{\sigma}_y$, and $\hat{\sigma}_z$.
Thereby, the Larmor frequencies $\omega_k$ 
of the first term characterize the interaction strengths
of these spins with an external inhomogeneous magnetic field so that these spins can be addressed
individually. The second term of the 
Hamiltonian \eqref{eq:sysham} describes 
the dipole-dipole interaction of the nuclear spins with
the coupling strength $J_{kl}$ between spins $k$ and $l$ being
inversely proportional to the cubic power of their distance.

If these nuclear spins are used as qubits of a quantum memory, for example, one has to protect them
against the perturbing influence of the interaction Hamiltonian \eqref{eq:sysham}. In the framework of a deterministic
decoupling scheme \cite{SV2005} this may be achieved by
an appropriate sequence of fast electromagnetic pulses.
Thereby, a series of fast global $\pi$-pulses is applied in order to suppress the Zeeman term $\hat{H}_Z$ while leaving the
dipole-dipole coupling term $\hat{H}_D$ invariant. 
This latter term can be suppressed by the well known WHH-4 pulse sequence  \cite{WHH68}.
This sequence consists of four fast
$\pi/2$-pulses applied at times $\tau$, $2\tau$, $4\tau$ and $5\tau$. Thus, the resulting
unitary time evolution after this pulse sequence, i.\,e. at time $T_c \equiv 6\tau$, is given by
\begin{equation}
\begin{split}
\hat{U}(t = 6\tau) &= \hat{U}^{\dagger}_1(-\tau)  \hat{U}_1(\tau)
\equiv  \prod_{j=1}^{6} e^{-i\hat{\tilde{H}}_j\tau/\hbar},\\
\hat{U}_1(\tau) &=  e^{-i\hat{H}\tau/\hbar} \hat{P}_{\overline{y}} e^{-i\hat{H}\tau/\hbar} \hat{P}_{{x}} e^{-i\hat{H}\tau/\hbar}
\end{split}
\end{equation}
with the interaction-picture (toggled) Hamiltonians
$\hat{\tilde{H}}_1 = \hat{\tilde{H}}_6 = \hat{H}_D$,
$\hat{\tilde{H}}_2 = \hat{\tilde{H}}_5 = \hat{P}_{\bar{x}} \hat{H}_D \hat{P}_x$
and $\hat{\tilde{H}}_3 = \hat{\tilde{H}}_4 = \hat{P}_{\bar{x}} \hat{P}_y \hat{H}_D \hat{P}_{\bar{y}} \hat{P}_x$.
Here, the unitary transformation $\hat{P}_x$ results from a fast global $\pi/2$-pulse witch generates rotations around the $x$-direction, i.\,e.
$\hat{P}_x = \bigotimes_{k=0}^{n_q-1} \exp ( -i \hat{\sigma}_x^k \pi/4 ) = \hat{P}^{\dagger}_{\overline{x}}$.
As a consequence, in zeroth-order average Hamiltonian theory (AHT) \cite{SV2005} the time-averaged Hamiltonian vanishes, i.\,e.
\begin{equation}
 \hat{\overline{H}}_D^{(0)} = \frac{1}{6} \sum_{j=1}^6 \hat{\tilde{H}}_j = 0.
\end{equation}
Due to the time reversal symmetry of the WHH-4 pulse sequence, i.\,e.
$\hat{\tilde{H}}(t)=\hat{\tilde{H}}(T_c-t)$,
in AHT all odd higher-order Hamiltonians vanish,
i.\,e. $\hat{\overline{H}}_D^{(2n+1)}=0$ for $n\in {\mathbb{N}}_0$.
\begin{figure*}
\includegraphics[scale=0.75]{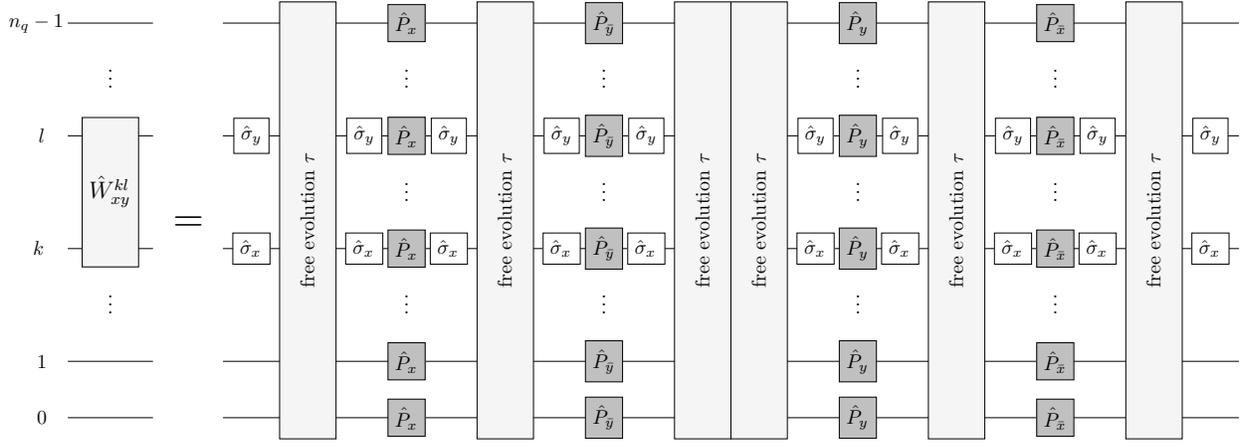}
\caption{Schematic representation of the unitary $\hat{W}_{xy}^{kl}$ quantum gate acting on qubits $k$ and $l$:
\textit{Free evolution} indicates time evolution according to the Hamiltonian $\hat{H}_D$ over a time interval of duration $\tau$.}
\label{fig:wxy}
\end{figure*}
\begin{figure*}
\includegraphics[scale=0.75]{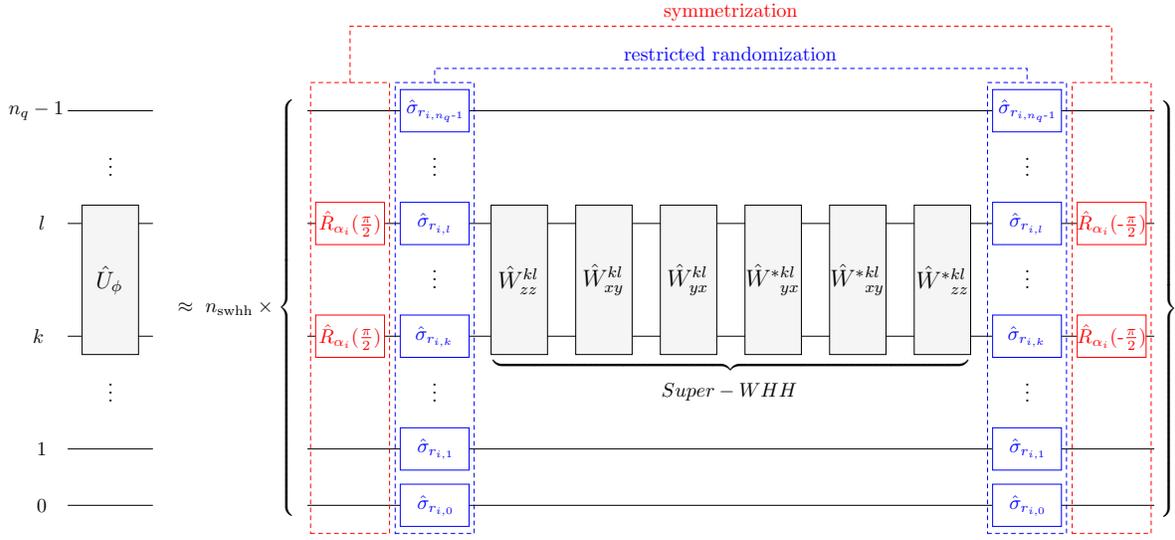}
\caption{Schematic representation of a conditional two-qubit gate ($\hat{U}_\phi$-gate) obtained by recoupling qubits $k$ and $l$ according to Eq. \eqref{eq:uphi}:
The $\hat{W}^*$ gates are obtained by reversing the order of the broadband pulses suppressing the Zeeman term.
Residual second-order terms of AHT can be eliminated by the restricted randomization step accomplished by random selective $\pi$-pulses $\hat{\sigma}_{r_{i,j}}$.
Thereby $r_{i,k}$ has to be equal to $r_{i,l}$ to ensure that the wanted gate action is not disturbed.
Still remaining terms are symmetrized by random $\pi/2$-pulses
$\hat{R}_{\alpha_i}(\pi/{2}) = \exp \bigl( -i \hat{\sigma}_{\alpha_i}\pi/4 \bigr)$,
$\alpha_i \in \{x,y,z\}$.
Either condition \eqref{eq:0bedingung} or condition \eqref{eq:2bedingung} has to be fulfilled depending on whether the original Super-WHH or the symmetrized Super-WHH sequence is used.
}
\label{fig:uphi}
\end{figure*}

If these nuclear spins are used as qubits of a quantum information processor one also has to 
implement universal sets of unitary quantum gates.
In particular, one needs to be able to implement two-qubit entanglement gates, such as controlled phase gates.
This can be accomplished
by recoupling qubits selectively with the help of a Super-WHH pulse sequence as proposed in Ref. \cite{YLMY04}.
Such a Super-WHH sequence recoupling qubits $k$ and $l$ consists of three WHH-4 sequences applied to the toggled Hamiltonians
$\hat{\tilde{H}}_{zz}^{kl} = \hat{\sigma}_z^k \hat{\sigma}_z^l \hat{H}_D \hat{\sigma}_z^k \hat{\sigma}_z^l$,
$\hat{\tilde{H}}_{xy}^{kl} = \hat{\sigma}_x^k \hat{\sigma}_y^l \hat{H}_D \hat{\sigma}_x^k \hat{\sigma}_y^l$, and
$\hat{\tilde{H}}_{yx}^{kl} = \hat{\sigma}_y^k \hat{\sigma}_x^l \hat{H}_D \hat{\sigma}_y^k \hat{\sigma}_x^l$, respectively.
Correspondingly, there are 18 time periods of duration $\tau$ during which the time evolution is described by
the double-toggled Hamiltonians
$\hat{\tilde{\tilde{H}}}_1 = \hat{\tilde{H}}_{zz}^{kl}$,
$\hat{\tilde{\tilde{H}}}_2 =\hat{P}_{\bar{x}}\hat{\tilde{H}}_{zz}^{kl}\hat{P}_x$, etc..
The appropriate
WHH-4 pulse sequence of the $\hat{\tilde{H}}_{xy}^{kl}$ Hamiltonian, for example,
is illustrated in Fig. \ref{fig:wxy}, where \textit{free evolution} denotes the time evolution according to the Hamiltonian $\hat{H}_D$
over a time interval of duration $\tau$.
The quantum gates resulting from these WHH-4 sequences are denoted by $\hat{W}_{xy}^{kl}$, $\hat{W}_{zz}^{kl}$, and $\hat{W}_{yx}^{kl}$, respectively .
The Super-WHH sequence is finally obtained from a combination of these latter quantum gates preceeded
by the corresponding time reversed sequence
(compare with the inner part of Fig. \ref{fig:uphi}).
As a consequence \cite{YLMY04}, this Super-WHH sequence yields 
the zeroth-order time-averaged Hamiltonian
\begin{equation}\label{eq:hr}
\hat{\overline{H}}_D^{(0)} = \frac{1}{36} \sum_{j=1}^{36} \hat{\tilde{\tilde{H}}}_j =
J^{(0)}_{kl} \bigl( \hat{\sigma}_x^k\hat{ \sigma}_x^l +\hat{ \sigma}_y^k\hat{ \sigma}_y^l +\hat{ \sigma}_z^k\hat{ \sigma}_z^l \bigr)
\end{equation}
with the renormalized zeroth-order recoupling strength
$J^{(0)}_{kl} = (J_{kl}/4)\times(8/9)$.
Due to the time reversal symmetry of the Super-WHH sequence, in AHT all odd-valued higher order recoupled Hamiltonians vanish, i.\,e.
i.\,e. $\hat{\overline{H}}_D^{(2n+1)} =0$ for $n\in {\mathbb{N}}_0$ .

With the help of the zeroth-order recoupled Hamiltonian $\hat{\overline{H}}_D^{(0)}$ of Eq. \eqref{eq:hr} one can approximate unitary two-qubit quantum gates of the form
\begin{equation}\label{eq:uphi}
\hat{U}_{\phi}^{kl}=\exp\bigl( -i
( \hat{\sigma}_x^k\hat{\sigma}_x^l + \hat{\sigma}_y^k\hat{\sigma}_y^l + \hat{\sigma}_z^k\hat{\sigma}_z^l )
\phi \bigr).
\end{equation}
Thereby, for a particular value of the phase $\phi$ one has to adjust the time $\tau$ between two successive pulses of a WHH-4 sequence and the number of times $n_\text{swhh}$ a Super-WHH sequence has to be applied according to the relation
\begin{equation}\label{eq:0bedingung}
J^{(0)}_{kl} \cdot n_\text{swhh} 36\tau/\hbar = \phi
\end{equation}
(compare with Fig. \ref{fig:uphi}).
However, because of the residual higher-order interactions which have not been canceled by the Super-WHH pulse sequence this implementation of a two-qubit quantum gate gate is only approximate.
\begin{widetext}
The error resulting from these residual higher-order
interactions is dominated by the second-order term of AHT which is generally given by \cite{nmrbook}
\begin{equation}
\hat{\overline{H}}^{(2)} = -\frac{1}{6T_c} \int_0^{T_c} dt_3 \int_0^{t_3} dt_2 \int_0^{t_2} dt_1
\Bigl([\hat{H}(t_3),[\hat{H}(t_2),\hat{H}(t_1)]] + [[\hat{H}(t_3),\hat{H}(t_2)],\hat{H}(t_1)] \Bigr)/\hbar^2
\end{equation}
for an arbitrary time-dependent Hamiltonian $\hat{H}(t)$.
In the case of the Super-WHH sequence we have
$\hat{H}(t)=\tilde{\tilde{H}}_j$ if $(j-1)\tau < t < j \tau$ and $T_c=36\tau$.
Therefore, according to AHT the lowest-order correction to the recoupled Hamiltonian of Eq. \eqref{eq:hr} is given by
\begin{equation}\label{eq:hd2}
\begin{split}
\hat{\overline{H}}_D^{(2)}= \sum_a^{\neq k,l} \Bigl[ &\hat{\sigma}_x^k \hat{\sigma}_x^l \Bigl( -322 J_{al}^2 J_{ak} +446 J_{ak}^2 J_{al}
+3628 J_{al} J_{ak} J_{kl} -2906 J_{ak}^2 J_{kl} -1370 J_{al}^2 J_{kl} \Bigr) + \\
 &\hat{\sigma}_y^k \hat{\sigma}_y^l \Bigl( +308 J_{al}^2 J_{ak} +308 J_{ak}^2 J_{al} +3208 J_{al} J_{ak} J_{kl}
 -2588 J_{ak}^2 J_{kl} -2588 J_{al}^2 J_{kl} \Bigr) +\\
 &\hat{\sigma}_z^k \hat{\sigma}_z^l \Bigl( +446 J_{al}^2 J_{ak} -322 J_{ak}^2 J_{al} +3580 J_{al} J_{ak} J_{kl}
 -1922 J_{ak}^2 J_{kl} -3458 J_{al}^2 J_{kl} \Bigr) \Bigr] \tau^2/(\hbar^2 \cdot 1728) + \dots . %(27 \cdot 4^3)
\end{split}
\end{equation}
Thereby, only terms of the form
$\hat{\sigma}_\alpha^k \hat{\sigma}_\beta^l \equiv \mathbf{1} \otimes \dots \otimes \mathbf{1} \otimes
\hat{\sigma}_\alpha^k \otimes \mathbf{1} \otimes \dots \otimes
\mathbf{1} \otimes \hat{\sigma}_\beta^l \otimes \mathbf{1} \otimes \dots \otimes \mathbf{1}$, ($\alpha,\beta \in \{x,y,z\}$)
are indicated as all other terms are irrelevant for our subsequent discussion.
\end{widetext}%
As a consequence, the Hamiltonian resulting from recoupling of qubits $k$ and $l$ by a Super-WHH sequence is of the form
\begin{equation}\label{eq:hrklp1}
\hat{H}_R^{kl'} =  J^{(0)}_{kl}
\bigl( \hat{\sigma}_x^k\hat{\sigma}_x^l + \hat{\sigma}_y^k\hat{\sigma}_y^l + \hat{\sigma}_z^k\hat{\sigma}_z^l \bigr)
+ \mathcal{O}\left(J (J\tau/\hbar)^2\right).
\end{equation}
To estimate the resulting error affecting the unitary gate $\hat{U}_\phi^{kl'}$ generated by $\hat{H}_R^{kl'}$ we have to multiply the $\mathcal{O}\left(J(J\tau/\hbar)^2\right)$
term of \eqref{eq:hrklp1} by the duration $n_\text{swhh}36\tau$ of this quantum gate.
Using condition \eqref{eq:0bedingung} we see that for a fixed value of the phase $\phi$ the resulting error depends on $n_\text{swhh}$ according to
$J^3 \tau^2 \cdot n_\text{swhh} 36 \tau/\hbar \sim \phi^3/(36 n_\text{swhh})^2$.
Therefore, the resulting contribution to the fidelity $f_{U_\phi'}$ of any quantum state $|\Psi\rangle$ is given by
\begin{equation}\label{eq:gatefid}
f_{U_\phi'} = \bra{\Psi}{} \hat{U}^\dagger_\phi \hat{U}_\phi' \ket{\Psi}{} = 1 - \mathcal{O}(\phi^6/(36 n_\text{swhh})^4).
\end{equation}

\subsection{Randomization and selective recoupling\label{sec:rec2}}
The residual interaction described by the Hamiltonian \eqref{eq:hd2} can be suppressed significantly
by embedding the recoupling scheme of Sec. \ref{sec:rec1} into a stochastic decoupling scheme,
such as the recently proposed PAREC scheme \cite{parec}.
For this purpose we choose at random an $n_q$-fold tensor product of Pauli-matrices
$\hat{\sigma}_{r_{i,0}} \otimes \hat{\sigma}_{r_{i,1}} \otimes \dots \otimes \hat{\sigma}_{r_{i,n_q-1}}$
($r \in \{0,x,y,z\}, \hat{\sigma}_0\equiv \hat{\mathbf{1}}$) and apply it before and after the $i$-th Super-WHH sequence.
This way each deterministic Super-WHH sequence is embedded within two statistically independent random Pauli operations.
In contrast to a usual dynamical decoupling scenario \cite{parec} 
in our case we have to choose the Pauli-matrices in such a way that
they leave the ideally recoupled gate Hamiltonian $\hat{\overline{H}}_D^{(0)}$ of Eq. \eqref{eq:hr} invariant.
This can be achieved by imposing the restriction
that the randomly chosen statistically independent Pauli spin operators have to be identical for qubits
$k$ and $l$ for each Super-WHH sequence, i.\,e. $\hat{\sigma}_{r_{i,k}}=\hat{\sigma}_{r_{i,l}}$.
This restriction assures that terms of the form $\hat{\sigma}_\alpha^k \hat{\sigma}_\alpha^l$ in $\hat{H}_R^{kl'}$ remain invariant (compare with Fig. \ref{fig:uphi}).
Since $\hat{\overline{H}}_D^{(2)}$ contains no terms of the form $\hat{\sigma}_\alpha^k \hat{\sigma}_\beta^l$ with $\alpha \neq \beta$
(compare with Eq. \eqref{eq:hd2}) the Pauli-matrices for qubits $k$ and $l$ can always be omitted, i.\,e. chosen to be the identity, $r_{i,k}=r_{i,l}=0$.

The only terms of the Hamiltonian $\hat{\overline{H}}_D^{(2)}$ 
which cannot be eliminated by this constrained randomization method are the ones containing terms of the form
$\hat{\sigma}_\alpha^k \hat{\sigma}_\alpha^l$ ($\alpha \in \{x,y,z\}$) which are shown in Eq. \eqref{eq:hd2}.
However, by an additional symmetrization these terms can be made rotationally invariant so that they can be cast into the form of Eq. \eqref{eq:hr}.
Thus, for a given value of $\phi$ these terms lead to a renormalization of the values of the required gate parameters
$\tau$ and $n_{swhh}$.
This rotational symmetrization can be achieved by selective $\pi/2$-pulses which induce unitary transformations of the form
$\hat{R}_\alpha (\varphi) = \exp(-i \hat{\sigma}_\alpha /2 \cdot \varphi )$ with $\alpha \in \{x,y,z\}$.
For this purpose one chooses 
one of the three unitary transformations $\hat{R}_{\alpha_i}^k(\pi/2) \hat{R}_{\alpha_i}^l(\pi/2)$ acting on qubits $k$ and $l$
randomly and applies it before and the corresponding inverse transformation after the $i$-th Super-WHH sequence
(compare with Fig. \ref{fig:uphi}). 
This way the coefficients of the $\hat{\sigma}_\alpha^k \hat{\sigma}_\alpha^l$-terms are permuted in the relevant toggled Hamiltonians.
As a consequence one obtains the statistically and rotationally averaged second-order contribution
\begin{equation}
\hat{\overline{\overline{H}}}_D^{(2)} =
\bigl( \hat{\sigma}_x^k \hat{\sigma}_x^l + \hat{\sigma}_y^k\hat{ \sigma}_y^l + \hat{\sigma}_z^k\hat{\sigma}_z^l \bigr) J^{(2)}_{kl} \tau^2/\hbar^2
\end{equation}
with
\begin{multline}
J^{(2)}_{kl} = \sum_a^{\neq k,l} \Bigl(  \frac{1}{12} ( J_{al}^2 J_{ak} + J_{ak}^2 J_{al} ) \\
+\frac{217}{108} J_{al} J_{ak} J_{kl} -\frac{103}{72} ( J_{ak}^2 J_{kl} +  J_{al}^2 J_{kl} ) \Bigr).
\end{multline}

By this combined randomization and symmetrization method the improved recoupled Hamiltonian 
\begin{multline}\label{eq:hrklp2}
\hat{H}_R^{kl''} = \Bigl(
J^{(0)}_{kl} + J^{(2)}_{kl} (\tau/\hbar)^2 + \mathcal{O}\bigl( J(J\tau/\hbar)^4 \bigr)
\Bigr) \cdot\\
\bigl( \hat{\sigma}_x^k \hat{\sigma}_x^l + \hat{\sigma}_y^k \hat{\sigma}_y^l + \hat{\sigma}_z^k \hat{\sigma}_z^l \bigr)
\end{multline}
is obtained.
In contrast to Eq. \eqref{eq:hrklp1}, now
the effective recoupling strength is renormalized and  the residual error is suppressed up to fourth-order in the small coupling parameter $J\tau/\hbar \ll 1$.
Thus, in order to implement a $\hat{U}_\phi$-gate, for example, we now have to choose the renormalized characteristic parameter $\tau'$ in such a way that the condition
\begin{equation}\label{eq:2bedingung}
\left( J^{(0)}_{kl} + J^{(2)}_{kl} (\tau'/\hbar)^2 \right) n_\text{swhh} 36 \tau' = \phi
\end{equation}
is fulfilled. As a result, in general
the required time of free evolution $\tau'$ depends on the chosen qubit pair $(k,l)$.

\section{Error analysis of a dynamically decoupled $\hat{U}_\phi$-gate\label{sec:pi4gate}}

In this section the stabilizing properties of selective recoupling by the embedded dynamical decoupling method of Sec. \ref{sec:rec2} 
is investigated for a unitary phase gate as described by Eq. \eqref{eq:uphi}.
As shown in Eq. \eqref{eq:gatefid}
the fidelity of a unitary phase gate $\hat{U}_\phi$ which is realized by recoupling of qubits $k$ and $l$ with the help of
the average Hamiltonian of Eq. \eqref{eq:hrklp1}
deviates from unity by terms of the order of $\mathcal{O}\left(\phi^6/(36n_\text{swhh})^4\right)$.
Thereby,
$n_\text{swhh}$ and $\tau$ denote the number of required iterations of the Super-WHH sequence
and the time required for the intermediate free evolution
as determined by relation \eqref{eq:0bedingung}, respectively.

In order to estimate the improvement achievable with the help of the embedded recoupling scheme
of Sec. \ref{sec:rec2} we start from the observation that
in the case of a time independent Hamiltonian $\hat{H}$
a rigorous lower bound of the mean fidelity ${\mathbb E}f (T)$
of a stochastic
dynamical decoupling scheme is given by \cite{VK05,combi}
\begin{equation}\label{bound}
\mathbb{E}f(T) \geq 1 -\Vert \hat{H}\Vert^2 \Delta t T/\hbar^2
\end{equation}
for $\Vert\hat{H}\Vert^2\Delta tT/\hbar^2\ll 1$.
Thereby, the norm of the Hamiltonian 
$\Vert \hat{H}\Vert$ is defined by its largest eigenvalue, $\Delta t$ is the time interval
between successively applied uncorrelated random pulses,
$\mathbb{E}$ denotes statistical averaging,
and $T$ is the interaction time.
This result can be used to estimate the mean
error bound of the fidelity of the corresponding embedded decoupling scheme in which the 
recoupling procedure leading to
the Hamiltonian \eqref{eq:hrklp1}
is embedded into statistically independent random Pauli operations. For this purpose we make the replacements
$\mathcal{O} \bigl(\Vert \hat{H} \Vert\bigr) \rightarrow J(J\tau/\hbar)^2$,
$\Delta t \rightarrow 36\tau$,
$T \rightarrow n_\text{swhh}36\tau$.
From relation \eqref{eq:0bedingung} we obtain the mean error estimate of the fidelity to be of the order of
$\mathcal{O}\left(\phi^6/ (36^4 n_\text{swhh}^5)\right)$.

\begin{figure}
 \centering
 \psset{unit=0.75cm}
 \begin{psmatrix}[colsep=1.8,rowsep=1.8,mnode=circle]
  [name=N3] \large 3 & [name=N2] \large 2 & [name=N1] \large 1 & [name=N0] \large 0
 \end{psmatrix}
 %wag Verbind.
 \ncline[]{N0}{N1}\naput{$J$}\ncline[]{N0}{N1}
 \ncline[]{N1}{N2}\naput{$J$}\ncline[]{N1}{N2}
 \ncline[]{N2}{N3}\naput{$J$}\ncline[]{N2}{N3}

 \vspace{2mm}
%GNUPLOT: LaTeX picture with Postscript
\begin{picture}(0,0)%
\includegraphics{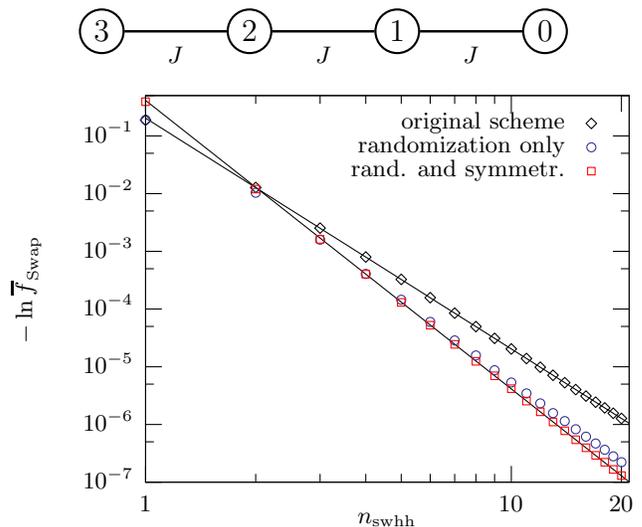}%
\end{picture}%
\begingroup
\setlength{\unitlength}{0.0200bp}%
\begin{picture}(14400,8640)(0,0)%
\put(2475,900){\makebox(0,0)[r]{\strut{}$10^{-7}$}}%
\put(2475,1988){\makebox(0,0)[r]{\strut{}$10^{-6}$}}%
\put(2475,3076){\makebox(0,0)[r]{\strut{}$10^{-5}$}}%
\put(2475,4165){\makebox(0,0)[r]{\strut{}$10^{-4}$}}%
\put(2475,5253){\makebox(0,0)[r]{\strut{}$10^{-3}$}}%
\put(2475,6341){\makebox(0,0)[r]{\strut{}$10^{-2}$}}%
\put(2475,7429){\makebox(0,0)[r]{\strut{}$10^{-1}$}}%
\put(11666,450){\makebox(0,0){\strut{}20}}%
\put(9591,450){\makebox(0,0){\strut{}10}}%
\put(2700,450){\makebox(0,0){\strut{}1}}%
\put(450,4545){\rotatebox{90}{\makebox(0,0){\strut{}$-\ln \overline{f}_\text{Swap}$}}}%
\put(7256,225){\makebox(0,0){\strut{}$n_\text{swhh}$}}%
\put(10612,7665){\makebox(0,0)[r]{\strut{}original scheme}}%
\put(10612,7215){\makebox(0,0)[r]{\strut{}randomization only}}%
\put(10612,6765){\makebox(0,0)[r]{\strut{}rand. and symmetr.}}%
\end{picture}%
\endgroup
%\endinput
 \caption{The fidelity (\textit{bottom}) of the $\hat{U}_{\pi/4}^{12}$-gate on a linear four-qubit chain (\textit{top})
as a function of the number of repetitions $n_\text{swhh}$ of the Super-WHH scheme:
original Super-WHH sequence (diamonds), unsymmetrized embedded
scheme (circles), and symmetrized embedded scheme with adapted pulse interval $\tau'$ according to Eq. \eqref{eq:2bedingung} (squares).
The solid lines represent the fitting functions $\exp(-c/n_\text{swhh}^4)$ and $\exp(-c_{rs}/n_\text{swhh}^5)$ with
$c = 0.2$ and $c_{rs} = 0.41$.}
\label{fig:pi4gate}
\end{figure}

In Fig. \ref{fig:pi4gate} (\textit{bottom}) the mean fidelity $\overline{f}_{{\rm Swap}}$
of a unitary $\hat{U}_{\pi/4}$-gate and its dependence on the number of performed Super-WHH sequences
$n_\text{swhh}$ is depicted. In these numerical simulations this unitary quantum gate is realized
by recoupling of the two central qubits of a linear four-qubit chain.
Apart from an irrelevant global phase
this unitary $\hat{U}_{\pi/4}$-gate is nothing but a \qgswap{}-gate (compare with Fig. \ref{fig:gates}).
The mean fidelity of Fig. \ref{fig:pi4gate}  was obtained by averaging over all $2^4$ orthonormal initial states of the computational basis.
The statistical averaging was performed over 100 runs with statistically independent realizations of the random pulses involved.
The coupling strength is assumed to be constant for adjacent qubits and to be vanishing between all other qubits
(compare with Fig. \ref{fig:pi4gate} (\textit{top}).
Fig. \ref{fig:pi4gate} (\textit{bottom}) demonstrates that
the fidelity (diamonds) resulting from non-randomized Super-WHH pulse sequences
can be fitted well by a function of the form $\exp(-c/n_\text{swhh}^4)$ with $c \approx 0.20$.
This is consistent with the simple estimate \eqref{eq:gatefid}.
Using a recoupling scheme based on the symmetrized embedded procedure discussed in Sec. \ref{sec:rec2} 
while choosing $\tau$ according to the modified condition \eqref{eq:2bedingung},
we notice that the resulting fidelity (squares) is fitted well by a function of the form $\exp(-c_{rs}/n_\text{swhh}^5)$ with $c_{rs} \approx 0.41$.
If symmetrization is omitted an intermediate behavior is obtained (circles).

\section{Numerical simulation of a quantum algorithm\label{sec:saw}}
In this section
the question is explored how much can be  gained by stabilizing an iterative quantum algorithm by the embedded recoupling scheme
of Sec. \ref{sec:rec2}.

\subsection{Quantum computation with the $\hat{U}_\phi$ gate\label{sec:qc}}
For purposes of quantum computation 
one needs to know how to perform two-qubit entanglement gates,
such as the controlled-not-gate (\qcnot-gate) or the controlled-phase-gate (\qgcp{}{\varphi}-gate), on the
basis of the recoupled
Hamiltonian $\hat{H}_R^{kl'}$. Definitely,
\begin{figure*}
% \hfill\includegraphics[scale=0.82]{g-swappic}%
 \hfill\includegraphics[scale=0.82]{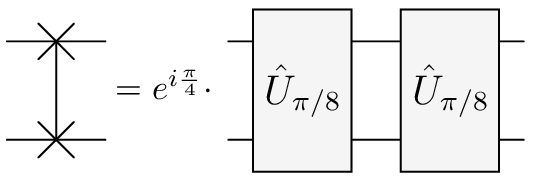}%
% \hfill\includegraphics[scale=0.82]{g-cnotpic}\hspace*{\fill}%
 \hfill\includegraphics[scale=0.82]{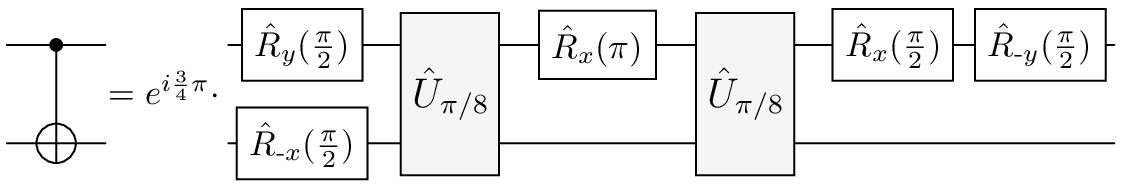}\hspace*{\fill}%

 \vspace{1cm}
\includegraphics[scale=0.82]{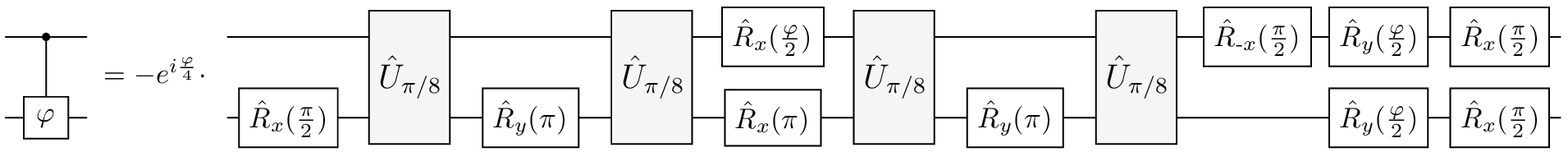}
 \caption{Quantum gates of the \qgswap{}, \qcnot{}{}, and controlled-phase gate (\qgcp{}{\varphi}):
$\hat{R}_{\pm\alpha} (\varphi) = \exp(\mp i \hat{\sigma}_\alpha /2 \cdot \varphi )$, $\alpha \in \{x,y,z\}$.}
 \label{fig:gates}
\end{figure*}
such quantum gates can be performed only between qubits $k$ and $l$ which are coupled, i.\,e. for which $J_{kl} \neq 0$.
Therfore, in order to be able to
entangle any two qubits of a quantum computer it is necessary to 
swap qubit pairs with vanishing coupling constants to neighboring positions.
Fortunately, such a unitary swapping gate can be realized easily by the unitary phase gate of Eq. \eqref{eq:uphi}
because
$\qgswap{kl}\equiv\hat{U}_{\pi/4}^{kl}$.
Throughout the rest of this paper we will use the quantum phase gate $\hat{U}_{\pi/8}^{kl}$
as a basic building block for all two-qubit quantum gates. Thus, the quantum 
\qgswap{kl}-gate consists of the repeated application of two such gates. 
For the realization of other two-qubit quantum gates repeated applications of this $\hat{U}_{\pi/8}$-gate in
combination with single-qubit gates are required.
In Fig. \ref{fig:gates} basic gate decompositions are depicted for the \qcnot-gate, the \qgcp{}{\varphi}, and for the \qgswap{}-gate.
These decompositions will be used in the next section for the simulation of a quantum algorithm.
The $\hat{U}_{\pi/8}$-gate itself can be generated approximately either by repeated application of the original or of the
randomized
Super-WHH recoupling sequence using either condition \eqref{eq:0bedingung} or relation \eqref{eq:2bedingung} for the 
determination of the free evolution time $\tau$ between successive fast pulses.

\subsection{Lattice model of a quantum computer\label{sec:saw2}}
For the subsequent numerical simulations of a quantum algorithm we consider a quantum information processor
consisting of
$n_q = 9$ qubits which are arranged on a lattice
as indicated in Fig. \ref{fig:qubits}.
\begin{figure}
\begin{minipage}[t]{.6\columnwidth}
 \vspace{0pt}
 \psset{unit=0.75cm}
 \begin{psmatrix}[colsep=1.8,rowsep=1.8,mnode=circle]
  [name=N6] \large 6 & [name=N7] \large 7 & [name=N8] \large 8 \\[0pt]
  [name=N3] \large 3 & [name=N4] \large 4 & [name=N5] \large 5 \\[0pt]
  [name=N0] \large 0 & [name=N1] \large 1 & [name=N2] \large 2
 \end{psmatrix}
 %wag dann senkr. Verbind.
 \ncline[]{N0}{N1}\naput{$J$}\ncline[]{N1}{N2}
 \ncline[]{N3}{N4}\ncline[]{N4}{N5}
 \ncline[]{N6}{N7}\ncline[]{N7}{N8}
 \ncline[]{N0}{N3}\naput{$J$}\ncline[]{N1}{N2}\ncline[]{N3}{N6}
 \ncline[]{N1}{N4}\ncline[]{N4}{N7}
 \ncline[]{N2}{N5}\ncline[]{N5}{N8}
 %Die diagonalen:
 \psset{nrot=:U}
 \ncline[linecolor=gray]{N0}{N4}\ncline[linecolor=gray]{N3}{N1}\naput{$J\cdot 2^{-3/2}$}
 \ncline[linecolor=gray]{N1}{N5}\naput{$J\cdot 2^{-3/2}$}\ncline[linecolor=gray]{N4}{N2}
 \ncline[linecolor=gray]{N3}{N7}\ncline[linecolor=gray]{N6}{N4}
 \ncline[linecolor=gray]{N4}{N8}\ncline[linecolor=gray]{N7}{N5}
\end{minipage}%
\begin{minipage}[t]{.4\columnwidth}
 \vspace{0pt}\vspace{0.5cm}
 \begin{tabular}{c|c}
  $(k,l)$ & $ J^{(2)}_{kl} / J^3$ \\
  \hline
  \hline
  $\{(0,1),(1,2),$ & $-\frac{923}{192}$ \\
  $\hphantom{\{}(6,7),(7,8),$ & $+\frac{113}{54 \sqrt{2}}$\\
  $\hphantom{\{}(0,3),(3,6),$ & \\
  $\hphantom{\{}(2,5),(5,8)\}$ & \\
  \hline
  $\{(1,4),(4,7),$ & $-\frac{2357}{288}$\\
  $\hphantom{\{}(3,4),(4,5)\}$  & $+\frac{113}{27 \sqrt{2}}$ \\
 \end{tabular}
\end{minipage}
\caption{\textit{Left}: 
The qubits of the
nine-qubit quantum information processor
are arranged on a lattice.
The lines connecting qubits $i$ and $j$ indicate the values of the coupling constants $J_{ij}$;
\textit{right}: The table shows the two different values of the coupling constants $J^{(2)}_{kl}$ for each qubit pair $(k,l)$.}
 \label{fig:qubits}
\end{figure}
The coupling constants of vertical or horizontal qubit pairs are assumed to be equal while the coupling constants of diagonal neighbors are smaller by a factor of $2^{-3/2}$ due to the larger distance between them.
Non-neighboring qubits are assumed to be uncoupled.
According to relation \eqref{eq:2bedingung} this implies that in the embedded recoupling scheme two different time intervals $\tau$ are required for the free evolutions.
The values of the coupling strengths $J^{(2)}_{kl}$ for the 9-qubit lattice used in our subsequent simulation are apparent from the table of Fig. \ref{fig:qubits}.

In the following it is assumed that a quantum algorithm is performed on this quantum information processor according to the following rules:
\begin{enumerate}
\item Single-qubit gates are performed instantaneously and perfectly.
\item Two-qubit gates between vertical or horizontal neighboring qubits are performed by repeated applications of the unitary
$\hat{U}_{\pi/8}$-gate of Sec. \ref{sec:qc} in combination with single-qubit gates as illustrated in Fig. \ref{fig:gates}.
The $\hat{U}_{\pi/8}$-gate itself is generated by applying Super-WHH sequences $n_\text{swhh}$ times as indicated in Fig. \ref{fig:uphi}.
\item If the target qubits of a two-qubit gate are not vertical or horizontal neighbors
they are moved into such positions by applying
a sequence of \qgswap{}-gates according to the following simple strategy
\footnote{A better but more complicated strategy would be to minimize the number of \qgswap{}-gates.
Note that due to the simple strategy used in this paper the first few iterations of a quantum algorithm take different amounts
of computation time because the initial positions of the logical qubits are varying and so does the number of \qgswap{}-gates.}:
If the vertical position of the qubits is the same, move the lower qubit to the upper one.
Otherwise, move the lower one to the same horizontal position and afterwards move the left one as far as necessary to the right.
\item A Super-WHH sequence is always applied in such a way that the qubit whose physical position has the smaller label
(compare with Fig. \ref{fig:qubits}) is  qubit $k$ in $\hat{W}_{xy}^{kl}$, i.\,e. it is transformed by the $\hat{\sigma}_x$ transformations.
The gate sequence of the \qgcp{}{\varphi}-gate (compare with Fig. \ref{fig:gates}) is applied in such a way that the first single-qubit gate
is applied always to the qubit with the smaller label.
\end{enumerate}

\subsection{The Quantum Algorithm}
In order to investigate the stabilizing properties of the embedded recoupling scheme the quantum algorithm of the quantum sawtooth map \cite{shep129} is simulated according to the rules of Sec. \ref{sec:saw2}.
One iteration of the quantum sawtooth map transforms an initial $n_q$-qubit quantum state $|\Psi\rangle$ to the quantum state
\begin{equation}
|\Psi\rangle' = e^{-iT\hat{p}^2/2}e^{-ik\hat{V}(\hat{\theta})/2}|\Psi\rangle
\end{equation}
with the sawtooth potential $V(\theta) = (\theta - \pi)^2$ ($0\leq \theta <2\pi$) and the (dimensionless) momentum operator $\hat{p}$.
Initially the nine-qubit quantum information processor is prepared in the momentum eigenstate $|\Psi\rangle = |100110011\rangle$.
The (dimensionless) parameters of the sawtooth map are assumed to have the same values as in the previous simulations of Ref. \cite{shep152}, i.\,e. $T=2\pi/2^{n_q}$ and $K \equiv kT = -0.5$.
Therefore, in Husimi-distributions, such as the ones presented in Fig. \ref{fig:husimi}, the dynamics of the sawtooth map is restricted to a phase-space cell of size $2\pi \times 2\pi$ and its corresponding classical dynamics are integrable.
In these Husimi-distributions the initial state corresponds to a horizontal line slightly above the middle.

Our gate decomposition of the quantum algorithm of this sawtooth map
consists of $n_g=2 n_q^2 + 2n_q$ quantum gates. It differs slightly from the gate decomposition of Ref. \cite{shep129}
since we are using the \qgcp{}{\varphi}-gate instead of the four-phases phase gate of Ref. \cite{shep129}.
In particular, $[2\cdot n_q(n_q+1)/2]$ quantum gates originate from the two quantum Fourier transforms
after which the inversion of the qubit positions is taken care of by relabeling instead of swapping.

\subsection{Numerical Results}
\begin{figure}
 \includegraphics[scale=0.82]{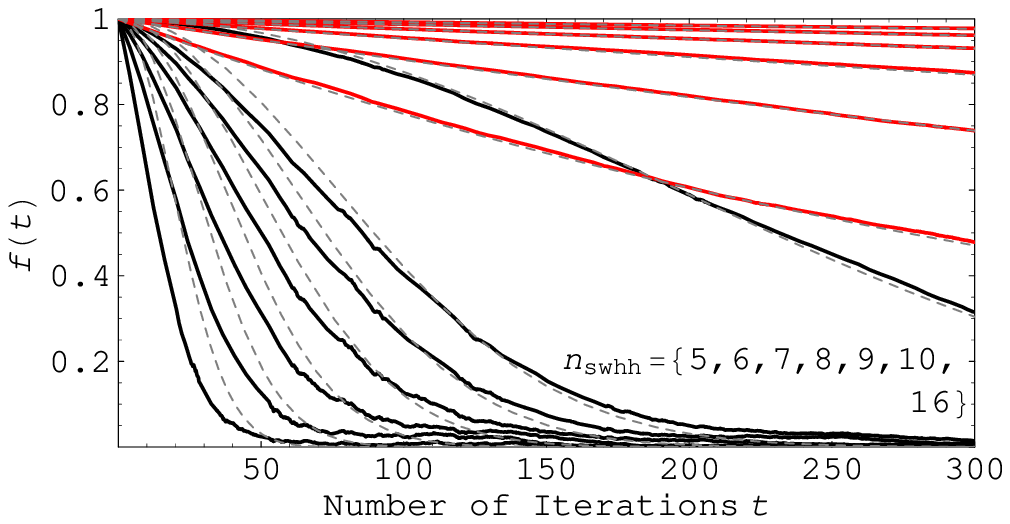}
 \vspace{2mm}

\includegraphics[scale=0.82]{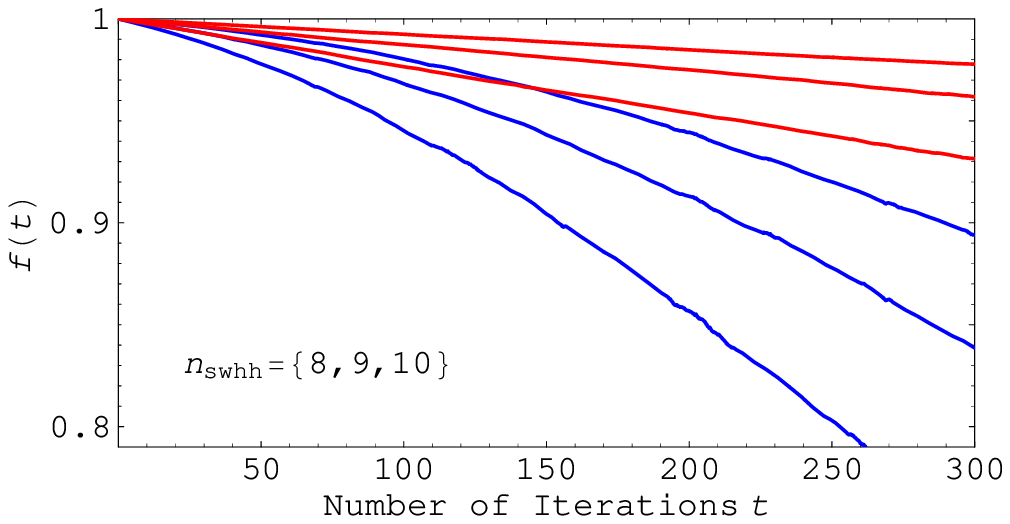}
\caption{
\textit{Upper plot:}
Fidelity plots of the quantum sawtooth map implemented with the original recoupling scheme of \citeauthor{YLMY04}\cite{YLMY04} (black)
and the corresponding plots of the symmetrized embedded recoupling scheme (red):
Dashed curves show the fidelity estimations according to Eqs. \eqref{eq:fidohne} and \eqref{eq:fidmit}.
\textit{Lower plot:}
Fidelity plots of the restricted randomized (blue) and of the symmetrized (red) embedded recoupling scheme. 
}
 \label{fig:fid}
\end{figure}
\begin{figure}
 \begin{minipage}[b]{0.5\columnwidth}
  \vspace{0pt}
  \centering
  \includegraphics[scale=0.7]{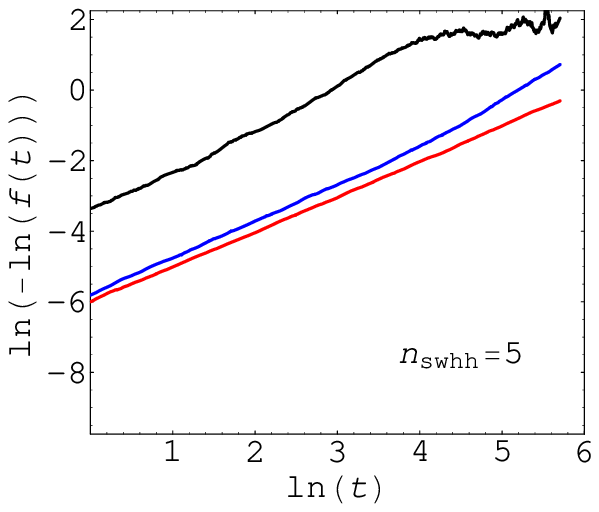}
 \end{minipage}%
 \begin{minipage}[b]{0.5\columnwidth}
  \vspace{0pt}
  \centering
  \includegraphics[scale=0.7]{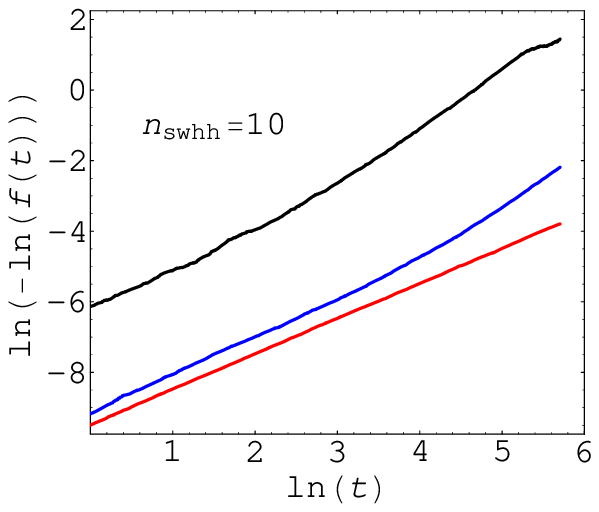}
 \end{minipage}%
  \caption{Logarithmic fidelity plots of the quantum sawtooth map with $n_\text{swhh}=5$ \textit{(left)} and $n_\text{swhh}=10$ \textit{(right)}:
the original \textit{\citeauthor{YLMY04}}\cite{YLMY04}  scheme (black),
the restricted but unsymmetrized embedded approach (blue),
and the restricted and symmetrized embedded scheme (red).
}\label{fig:lnln}
\end{figure}
\begin{figure}
 \begin{minipage}[b]{0.333\columnwidth}
  \vspace{0pt}
  \centering
  \includegraphics[scale=0.315]{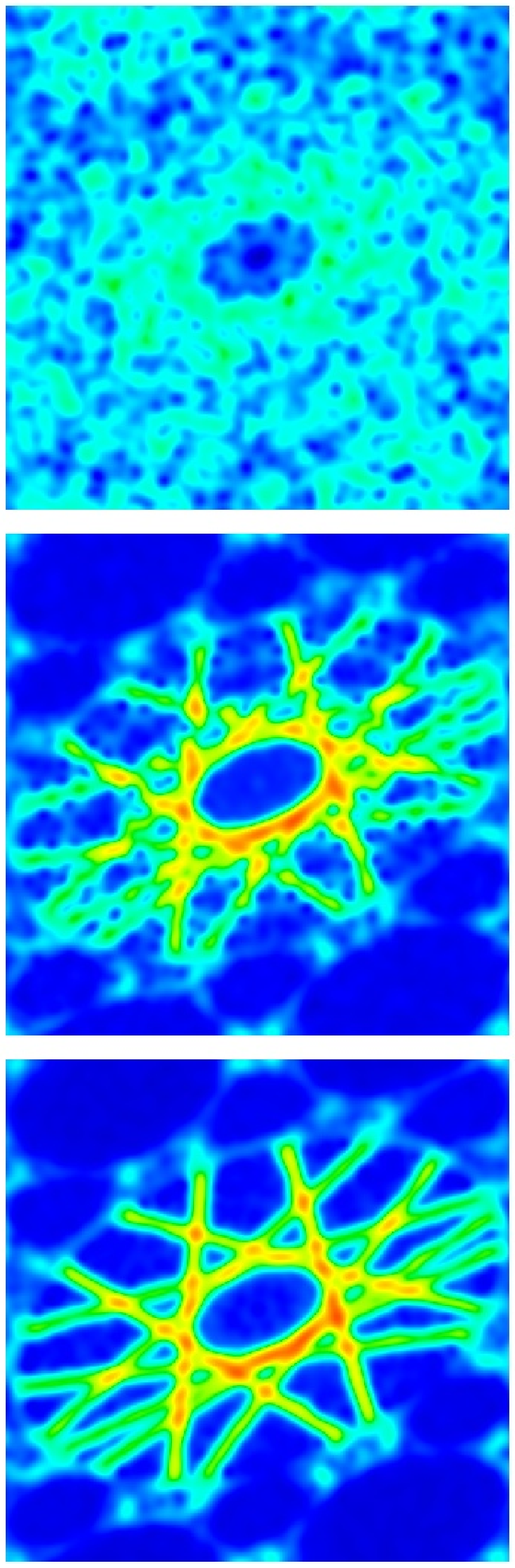}
  $n_\text{swhh}=6$
 \end{minipage}%
 \begin{minipage}[b]{0.333\columnwidth}
  \vspace{0pt}
  \centering
  \includegraphics[scale=0.315]{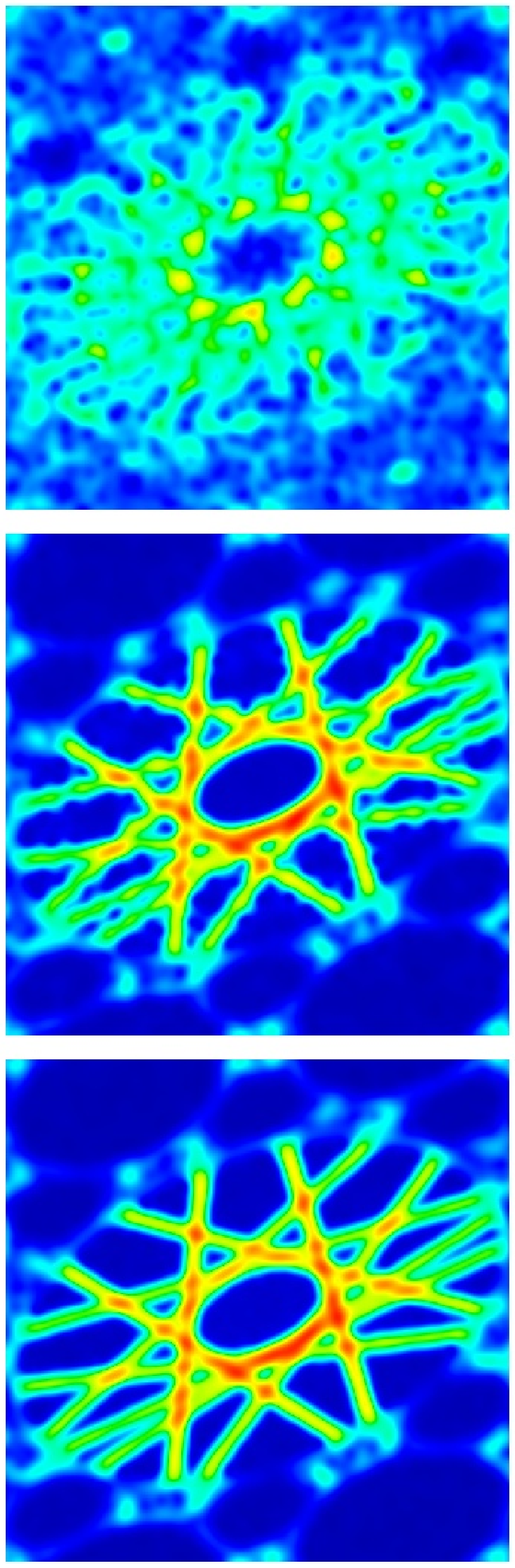}
  $n_\text{swhh}=8$
 \end{minipage}%
 \begin{minipage}[b]{0.333\columnwidth}
  \vspace{0pt}
  \centering
  \includegraphics[scale=0.315]{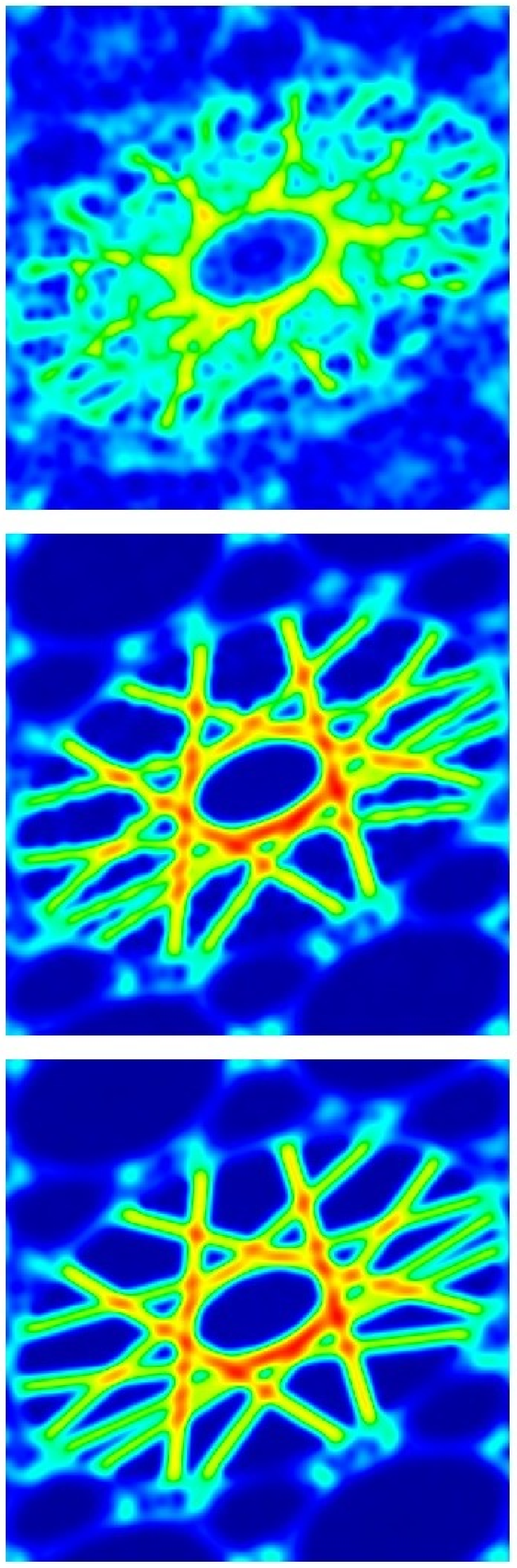}
  $n_\text{swhh}=10$
 \end{minipage}
 \caption{Husimi-distributions of the quantum states resulting from the quantum sawtooth map:
\textit{(Upper row)} The original scheme of \citeauthor{YLMY04}\cite{YLMY04},
\textit{(Middle row)} the embedded but unsymmetrized approach,
\textit{(Lower row)} the embedded and symmetrized scheme.
The $\hat{U}_{\pi/8}$-gates used in the computations consist of $n_\text{swhh} = \{6,8,10\}$ Super-WHH sequences (from left to right).
These distributions are averaged over $290 \leq t \leq 299$ numbers $t$ of iterations of the sawtooth map.
}
 \label{fig:husimi}
\end{figure}
In Figs. \ref{fig:fid},\ref{fig:lnln}, and \ref{fig:husimi} results of our numerical simulations of the fidelity
\begin{equation}
f(t) = \vert  \braket{\Psi(t)}{\Psi_\text{ideal}(t)}  \vert^2
\end{equation}
are presented for
different numbers of repetitions $n_\text{swhh} \in \{5,6,\dots,10,16\}$ of the Super-WHH sequences.
For each value of $n_\text{swhh}$ we calculated the fidelity
of the quantum state $|\Psi (t)\rangle$ of the quantum sawtooth map for up to $t=300$ iterations as well as the corresponding
Husimi-distributions.

The quadratic-in-time fidelity decay of the non-randomized recoupling scheme is clearly apparent from 
Figs. \ref{fig:fid} and \ref{fig:lnln}. (The corresponding fidelities are plotted in black color).
This decay is caused by the coherent accumulation of errors due to the second-order AHT-term of the Super-WHH sequences involved in the realization of the unitary $\hat{U}_{\pi/8}$-gate \cite{shep144}.
The $t$-dependence of the fidelity can be fitted by the function
\begin{equation}\label{eq:fidohne}
f(t) = \exp \bigl( - c \cdot t^2 / n_\text{swhh}^4 \bigr)
\end{equation}
with $c \approx 0.87$ (compare with the seven lowest dashed lines of the upper picture of Fig. \ref{fig:fid}). 
According to Ref. \cite{shep144} there should also be a linear contribution in the exponent of \eqref{eq:fidohne} which dominates the
fidelity decay for small numbers of iterations. Neglecting this linear
contribution is the reason
for the slightly imperfect overlap of our fitted fidelities with the corresponding numerical results.

Simulations based on the restricted embedded recoupling scheme without symmetrization are shown in Fig. \ref{fig:fid}
(lower part, blue plots) and Fig. \ref{fig:lnln} (blue plots).
The fidelity decay is suppressed significantly but on the time scale of these plots
it is still quadratic in time. This originates from the fact that 
terms of the Hamiltonian of Eq. \eqref{eq:hd2}
of the form $\hat{\sigma}_\alpha^k \hat{\sigma}_\alpha^l$, $\alpha \in \{x,y,z\}$, are not eliminated by the
PAREC method without symmetrization.

Using the randomized and symmetrized Super-WHH sequence together with the appropriately chosen free evolution times (compare with Eq. \eqref{eq:2bedingung}) it is possible to get an almost linear-in-time fidelity decay at least on time scales where errors of the order of $\mathcal{O}\bigl( J(J\tau/\hbar)^4 \bigr)$ are negligible (compare with Figs. \ref{fig:fid} and \ref{fig:lnln} (red plots)).
In these cases the fidelity decay can be fitted by the function
\begin{equation}\label{eq:fidmit}
f(t) = \exp \bigl( - c_{rs} \cdot t / n_\text{swhh}^5 \bigr)
\end{equation}
with $c_{rs} \approx 7.85$ (compare with the six upper dashed lines of the upper picture of Fig. \ref{fig:fid} which are almost indistinguishable from the corresponding full curves).

\section{Conclusions\label{sec:conclude}}
We showed how a selective recoupling scheme can be embedded into a stochastic decoupling scheme in such a way
that a particularly selected coupling is achieved and that, in addition, the coherent accumulation of higher-order errors is suppressed significantly.
Even if computation times of a quantum information processor are so long that the residual higher-order interaction term of Eq. \eqref{eq:hrklp2} of the order or $\mathcal{O}\bigl( J(J\tau/\hbar)^4 \bigr)$ is no longer negligible, it is possible to suppress also these errors significantly by a suitable adjustment of the free evolution time $\tau$ involved in the realization of the relevant two-qubit gates ($\hat{U}_\phi$-gates).
In generalization of the procedure discussed in Sec. \ref{sec:rec2} (compare with Eq. \eqref{eq:2bedingung}) this can be achieved either by explicitly calculating the fourth-order contribution of AHT and by solving the corresponding implicit
equation of fifth order for $\tau$ involving renormalized coupling strengths or, alternatively, adjusting the value of $\tau$ so that the resulting fidelity decay is as small as possible.

Basic properties of our embedded stabilization scheme were analyzed for a single two-qubit gate.
In particular, it was demonstrated that our proposed embedded symmetrized recoupling scheme results in an improvement of the scaling of the error of a swapping gate with $n_\text{swhh}^{-5}$ instead of $n_\text{swhh}^{-4}$.
Thereby, $n_\text{swhh}$ denotes the number of repetitions of an embedded Super-WHH sequence which are required for the realization of the phase gate.
Therefore, in our embedded recoupling scheme fewer numbers of repetitions of Super-WHH sequences are necessary for achieving a particular degree of error suppression.
Typically, this also implies fewer pulses which are required for performing a quantum computation with a particular error tolerance.
This aspect is apparent from the upper plot of Fig. \ref{fig:fid} where  at $t\approx 70$ iterations the fidelity of the original recoupling scheme with $n_\text{swhh}=16$ is the same as the one of the embedded symmetrized recoupling scheme with $n_\text{swhh}=6$.

Finally, we want to address briefly the additional phase shifts each selective pulse is accompanied by in any recoupling scheme.
These phase shifts (compare, e.\,g., with Eq. (24) of Ref. \cite{YLMY04}) originate from the transformation from the interaction picture
%needed for selective pulses
to the Schr\"odinger picture.
%needed for the recoupling scheme.
If necessary, such a phase shift can be eliminated by letting the system evolve freely after the application of a selective pulse for the same amount of time thereby embedding this additional time evolution between two broadband $\pi$ pulses.
This way an additional phase shift with the opposite sign is generated.

\begin{acknowledgments}
This work is supported by the EU projects EDIQIP and SECOQC.
O.~K. also acknowledges financial support by a Marie Curie fellowship
within the program `quantum information, computation, and complexity'
which took place at the Institut Henri Poincar\'e.
The authors thank this institution for its hospitality
and D. L. Shepelyansky for informative discussions.
\end{acknowledgments}

% Create the reference section using BibTeX:
%\bibliography{/home/okern/Documents/references}
\bibliography{references}

\end{document}